\documentclass[11pt,a4paper]{article}
\usepackage[utf8]{inputenc}
\usepackage{amsmath}
\usepackage{hyperref}
\usepackage{amsfonts}
\usepackage{amssymb}
\usepackage{graphicx}
\usepackage{authblk}
\usepackage{wrapfig}
\usepackage{caption}
\usepackage{color}
\usepackage{fullpage}
\usepackage{enumitem}
\usepackage{subcaption} 

\setlist{noitemsep}

\newcommand{\av}[1]{\langle #1 \rangle}

\newcommand{\bG}{\mathbb G}

\newcommand{\mink}{\mathbb M}
\newcommand{\bC}{\mathbf C} 
\newcommand{\bK}{{\mathbf K}} 
  
\newcommand{\bL}{\mathbf L}

\newcommand{\bl}{\mathbf l}
\newcommand{\bHH}{\mathbf H}

\newcommand{\tx}{\widetilde x}
\newcommand{\ty}{\widetilde y}
\newcommand{\tO}{\tilde \Omega}
\newcommand{\appropto}{\mathrel{\vcenter{
  \offinterlineskip\halign{\hfil$##$\cr
    \propto\cr\noalign{\kern2pt}\sim\cr\noalign{\kern-2pt}}}}}

\author[1]{\textbf{Nomaan X}}
\affil[1]{\textit{Raman Research Institute, C.V. Raman Avenue, Sadashivnagar, Bangalore 560 080, India}}
\author[2]{\textbf{Fay Dowker}}
\affil[2]{\textit{Blackett Laboratory, Imperial College, London, SW7 2AZ, UK}}
\author[1]{\textbf{Sumati Surya}}
\title{\textbf{Scalar Field Green  Functions\\
 on  Causal Sets}}
\date{}
\begin{document}
\baselineskip0.65cm
\parskip0.4cm
\maketitle


\renewcommand{\abstractname}{}
\begin{abstract}
  We examine the validity and scope of Johnston's models for scalar field retarded Green functions
  on causal sets in 2 and 4 dimensions.  As in the continuum, the massive Green function can be
  obtained from the massless one, and hence the key task in causal set theory is to first identify
  the massless Green function. We propose that the 2-d model provides a Green function for the
  massive scalar field on causal sets approximated by any topologically trivial 2 dimensional
  spacetime. We explicitly demonstrate that this is indeed the case in a Riemann normal
  neighbourhood. In 4-d the model can again be used to provide a Green function for the massive
  scalar field in a Riemann normal neighbourhood which we compare to Bunch and Parker's continuum
  Green function. We find that the same prescription can also be used for deSitter spacetime and the
  conformally flat patch of anti deSitter spacetime.  Our analysis  then allows us to suggest a generalisation of Johnston's
  model for the Green function  for a causal set approximated by 3 dimensional flat spacetime.
\end{abstract}


\section{Introduction} 

Understanding classical and quantum scalar field propagation on a fixed causal set is an important problem
in causal set quantum gravity~\cite{blms,daughton,johnston,johnstonthesis,dalemb,dalembgen}. 
Although ignoring back reaction and the quantum dynamics of the causal set background itself
means that the treatment of scalar field dynamics
will be inconsistent in some way, we can hope to learn something about causal set theory by studying this 
problem. Recent progress in defining scalar quantum field theory on a causal set puts 
great importance on the retarded Green function for the field on the causal set. Such a Green function
can be used to obtain the Feynman propagator, or equivalently the Wightman function, of 
a distinguished quantum state  on a causal set $C$, the \textit{Sorkin-Johnston
state}~\cite{johnstontwo,aas}. This could have potentially interesting phenomenological consequences
\cite{aas}.  Sorkin's related construction of a double path integral form of 
free scalar quantum field theory on a finite casual set is also based on the retarded Green function \cite{sorkincausalsetqft}. 

 Johnston found massive scalar field retarded Green functions, $K_m(x,x')$, for  causal sets approximated by $d=2$ and $d=4$
Minkowski spacetime \cite{johnston,johnstonthesis}. For each case, he used a ``hop-stop'' ansatz 
in which the Green function equals a sum over appropriately chosen causal trajectories between
the two arguments of the Green function, 
with a weight assigned for every hop between the elements
of the trajectory and another for every stop at an intervening element. Requiring that the
continuum limit of the mean of the causal set Green function over Poisson point process 
samplings -- so-called sprinklings -- of the Minkowski spacetime equals the 
continuum retarded Green function  then fixes these
weights. Extending the scope of the hop-stop ansatz to a larger class of 
causal sets would therefore allow us to study causal set quantum field theory
further.
  
We begin by describing Johnston's model in Section \ref{sec:model} and explain how it can be
motivated by a spacetime treatment. We will see that the key is to identify the appropriate retarded
Green function for the \textit{massless} field. This then leads to our proposed extensions of the
model in Section \ref{sec:generalisations}.  For $d=2$ we propose that Johnston's definition of
$K_m(x,x')$ can be used for a minimally coupled massive scalar field on a causal set approximated by
any topologically trivial spacetime. The proposal stems from the fact that the massless, minimally
coupled scalar field theory is conformally invariant.  In $d=2$, for non minimal coupling --
\textit{i.e} with arbitrary coupling to the Ricci scalar -- we show that the Johnston $K_m(x,x')$ is
an appropriate retarded Green function in an approximately flat Riemann Normal Neighbourhood, up to
corrections.  In $d=4$ we find that it is possible to extend the Minkowski spacetime prescription to
an RNN, as well as de Sitter spacetime and the conformally flat patch of anti de Sitter
spacetime.  In all cases, the comparison with the continuum fixes the hop-stop weights. Our results
are \textit{exact} for de Sitter and the globally hyperbolic patch of  anti de Sitter spacetime, i.e.,   the limit of the mean of the massless
causal set Green function is the conformally coupled massless Green function. 
In Section
\ref{sec:discussion} we use this framework to propose a construction of the retarded Green functions
on $d=3$
Minkowski spacetime. Throughout the paper we assume that the spacetimes we are considering are
globally hyperbolic.

\section{The Model}
\label{sec:model}

Consider the massless scalar retarded Green function $G_0(x,x')$ on a
globally hyperbolic  $d$ dimensional spacetime $(M, g)$:
\begin{equation} \label{greenfunc}
\Box_x G_0(x,x')=- \frac{1}{\sqrt{-g(x')} }\delta(x-x')\,.
\end{equation}
We can formally write down the massive retarded Green function,
$G_m$, satisfying
\begin{equation}
(\Box_x - m^2) G_m(x,x')= - \frac{1}{\sqrt{-g(x)} }\delta(x-x')\,,
\end{equation}
as a formal expansion
\begin{equation}
\label{conv}
G_m=G_0 - m^2\,G_0*G_0 + m^4 \,G_0*G_0*G_0+ \ldots = \sum_{k=0}^\infty(-m^2)^k \underbrace{G_0 * G_0* \ldots G_0}_{k+1}
\end{equation}
where
\begin{equation}
(A\ast B)(x,x')\equiv \int d^dx_1 \sqrt{-g(x_1)} A(x,x_1) B(x_1,x')\,.
\end{equation}
Note that if $G_0(x,x')$ is retarded (\textit{i.e.} only nonzero
if $x'$ is in the causal past of $x$) then so is $G_m(x,x')$. Also
note that since $G_0$ is retarded, the convolution integrals are over finite regions of spacetime, subsets
of the causal interval between $x$ and $x'$.
This relation can be reexpressed in the compact form
\begin{equation}
\label{cconv}
G_m=G_0 - m^2 G_0*G_m\,.
\end{equation}
Conversely  $G_0$ can be obtained  from  $G_m$ via
\begin{equation}
\label{oppconv}
G_0=\sum_{k=0}^\infty(m^2)^k \underbrace{G_m * G_m* \ldots G_m}_{k+1}
\end{equation}
and
\begin{equation}
G_0=G_m+m^2 G_m *G_0 = G_m + m^2  G_0*G_m\,.
\end{equation}
Once we have the massless retarded Green function, then, we can write down a formal series
for the massive retarded Green function.

Now suppose we have in hand a  massless retarded Green function analogue, $K_0(x,x')$, on a causal set
which is a sprinkling at density $\rho$ into the
$d$-dimensional spacetime. We can immediately propose a massive retarded Green function
$K_m(x,x')$ on that causal set via the replacement
\begin{equation}
\int \sqrt{-g(x)}\,d^d x \rightarrow  \rho^{-1} \sum_{\textrm{causal set elements} }\,,
\end{equation}
leading to
\begin{equation}
\label{convK}
K_m =  K_0- \frac{m^2}{\rho} K_0*K_0+\frac{m^4}{\rho^2} K_0*K_0*K_0 + \ldots = \sum_{k=0}^\infty\left(-\frac{m^2}{\rho}\right)^k \underbrace{K_0 * K_0* \ldots K_0}_{k+1}
\end{equation}
where now the convolutions have become finite sums over causal set elements in the finite order interval between
$x$ and $x'$ and the series terminates and is well-defined for each pair $x$ and $x'$.

We will now show that Johnston's hop-stop models for the massive retarded Green functions
on causal sets approximated by
2 and 4 dimensional Minkowski space are based on natural
causal set analogues
of the massless Green functions .

\subsection{$d=2$ Minkowski spacetime}

The massless retarded Green  function in $d=2$ Minkowski spacetime $\mink^2$ is
\begin{equation}
\label{masslesscont2d}
G^{(2)}_0(x,x')=  \frac{1}{2}\theta(x_0-x'_0)\theta(\tau^2(x,x'))
\end{equation}
where $\tau(x,x')$  is defined by
\begin{align}
\label{tau2d}
&\tau(x,x') =   \sqrt{  (x_0-x'_0)^2 - (x_1 - x'_1)^2 } \ \ \textrm{when} \ \ (x_0-x'_0)^2 \ge (x_1 - x'_1)^2 \nonumber \\
&\textrm{and}\nonumber \\
&\tau(x,x') = \,i \, \sqrt{ -  (x_0-x'_0)^2 + (x_1 - x'_1)^2  } \ \ \textrm{when} \ \  (x_0-x'_0)^2 < (x_1 - x'_1)^2  \,.
\end{align}
$\theta$ is the Heaviside step function.

Now consider, on a causal set, the \textit{causal matrix}
\begin{equation}
C_0(x,x') \equiv
\left\{
        \begin{array}{ll}
                1  & \mbox{if } x' \prec x \\
                0 & \mbox{} \mathrm{otherwise}
        \end{array}
\right.
\end{equation}
for all elements $x$ and $x'$ of the causal set, where by $\prec$ we mean causal precedence. 
The Poisson point process of sprinkling at density $\rho$ in 2 dimensional Minkowski spacetime
gives rise to a random variable, $\mathbf{C}_0(x,x')$
 for every two points, $x$ and $x'$, of Minkowski spacetime
via the addition of $x$ and $x'$ to the sprinkled causal set and the evaluation of $C_0(x,x')$ on that causal set.
Actually, in this case, the random variable takes the same value -- the expected value -- in each realisation.
 It was observed in  \cite{daughton} that this value is
\begin{equation}
\label{expC0}
\av{\bC_0(x,x')} =  2 G^{(2)}_0(x,x')\,.
\end{equation}
This leads to the proposal for a massless retarded Green function, $K^{(2)}_0(x,x')$,
on a $d=2$ flat sprinkled causal set:
\begin{equation}
\label{massless2d}
K^{(2)}_0(x,x')\equiv  \frac{1}{2}C_0(x,x').
\end{equation}
We define a massive Green function $K^{(2)}_m(x,x')$ on $C$ using this $K^{(2)}_0(x,x')$ and
(\ref{convK}).

Let us define a $k$-\textit{chain between} $x'$ \textit{and} $x$ in a causal set  $C$ as  a totally
ordered subset of $C$,  $\{x_1,x_2,\ldots, x_k \}$ such that $x'\prec x_1\prec
x_2\prec....x_{k-1}\prec x_k\prec x$. For $k\ge 1$, define $C_k(x,x')$ to be the number of $k$-chains between $x$ and $x'$ when $x'\prec x$ and zero when $x' \not\prec x$.
The $C_k$'s are powers of the causal matrix:
\begin{equation}
\label{numberchains}
C_k(x,x') = \underbrace{C_0 * C_0* \ldots C_0}_{k+1} (x,x')\,.
\end{equation}
This then gives
\begin{equation}
\label{discreteGF2d}
K^{(2)}_m(x,x') =  \sum\limits_{k=0}^\infty  \biggl(- \frac{m^2}{\rho}\biggr)^{k}\biggl(\frac{1}{2}\biggr)^{k+1} C_k(x,x')\,,
\end{equation}
where the sum is written as an infinite sum but terminates for each pair $x$ and $x'$.

For each two points $x$ and $x'$ of $\mink^2$  and each $k$ the random variable
$\bC_k(x,x')$ is $C_k(x,x')$ evaluated on a sprinkled causal set including $x$ and $x'$,
and hence we have the random variable $\bK^{(2)}_m(x,x')$:
 \begin{equation}
\label{discreteGF2drandom}
\bK^{(2)}_m(x,x') \equiv   \sum\limits_{k=0}^\infty  \biggl(-\frac{m^2}{\rho}\biggr)^{k}\biggl(\frac{1}{2}\biggr)^{k+1} \bC_k(x,x').
\end{equation}
 Its mean -- for any sprinkling density -- is equal to the continuum Green function since
\begin{equation}
\label{abundancechains}
 \av{\bC_k(x,x')} = \rho^k (\underbrace{\av{\bC_0}\ast \ldots \ast \av{\bC_0}}_{k+1}) (x,x')
\end{equation}
and so
\begin{align}
\av{\bK^{(2)}_m(x,x')} & = \sum\limits_{k=0}^{\infty} \biggl(-\frac{m^2}{\rho}\biggr)^{k}\biggl(\frac{1}{2}\biggr)^{k+1} \av{\bC_k(x,x')} \label{proof1}\\
& = \sum\limits_{k=0}^{\infty}(-m^2)^{k}\underbrace{G^{(2)}_0 * G^{(2)}_0* \ldots G^{(2)}_0}_{k+1} (x,x')\label{proof2}\\
&= G^{(2)}_m(x,x')\label{proof3}\,.
\end{align}

In \cite{johnston} $K^{(2)}_m(x,x')$ was expressed in terms of the hop and stop weights, $a$ and $b $ respectively:
\begin{equation}
\label{hopstop}
K^{(2)}_m(x,x') =  \sum\limits_{k=0}^\infty  a^{k+1} b^k C_k(x,x').
\end{equation}
This form was described by Johnston using a particle language as a sum over all chains
between $x$ and $x'$:
for each $k$-chain the hop between two
successive elements is assigned the weight $a$ and the stop at each intervening element between
$x$ and $x'$ is assigned the weight $b$.  Now we see that the weight $a=\frac{1}{2}$ is associated to
 each factor of $K^{(2)}_0$ -- from the relationship between $K^{(2)}_0$ and the
 causal matrix -- and the weight $b=-\frac{m^2}{\rho}$ to each convolution. In
 \cite{johnston} a momentum space calculation was used to find $b$, but as we have just seen the
spacetime formulation is sufficient to read off the value.

\subsection{$d=4$ Minkowski spacetime}

In $d=4$ Minkowski spacetime, $\mink^4$, the retarded Green  function for the massless field only
has support on the light cone:
\begin{equation}
\label{cont4d}
G^{(4)}_0(x,x')= \frac{1}{2\pi}\theta(x_0-x'_0) \delta( \tau^2(x,x'))\,,
\end{equation}
where
\begin{align}
\label{tau4d}
&\tau(x,x') =   \sqrt{  (x_0-x'_0)^2 - (x_1 - x'_1)^2 -   (x_2 - x'_2)^2 - (x_3 - x'_3)^2} \ \ \textrm{when} \nonumber \\
& \ \ \ \ \ \ \ \ \ \ \ \ (x_0-x'_0)^2 \ge (x_1 - x'_1)^2 + \dots +(x_{3} - x'_{3})^2 \nonumber \\
&\textrm{and}\nonumber \\
&\tau(x,x') = \, i \, \sqrt{ -   (x_0-x'_0)^2 + (x_1 - x'_1)^2 +  (x_2 - x'_2)^2 + (x_3 - x'_3)^2 } \ \ \textrm{when} \nonumber \\
& \ \ \ \ \ \ \ \ \ \ \ \ (x_0-x'_0)^2 < (x_1 - x'_1)^2 + \dots +(x_3 - x'_3)^2  \,.
\end{align}

The causal set analogue is proportional to the \textit{link matrix}
\begin{equation}
L_0(x,x'):=
\left\{
        \begin{array}{ll}
                1  & \mbox{if } x' \prec x\,\,\mathrm{and} \,\,|[x,x']|=0 \\
                0 & \mbox{} \mathrm{otherwise}
        \end{array}
\right.
\end{equation}
where the \textit{exclusive interval} is defined by $[x,x'] \equiv \{ z\in C \,|\, x'\prec z \prec x \}$.
When  $x'\prec x$ and $|[x,x']|=0$ this relation is called a \textit{link}. The expectation value
of the corresponding random variable $\bL_0(x,x')$ in a
Poisson sprinkling of density $\rho$ is
\begin{equation}
\label{linkexp}
\av{\bL_0(x,x')}=\theta(x_0-x'_0) \theta( \tau^2(x,x'))\exp(-\rho V(x,x')),
\end{equation}
where $V(x,x')$ is the volume of the spacetime interval $J^-(x) \cap
J^+(x')$. Here $J^{+}(x)$ and $J^{-}(x)$ denote \footnote{see Wald for
  example} the causal future and past of $x$, respectively. In $\mink^4$,
$V(x,x')= \frac{\pi}{24}\tau^4(x,x')$, so that
\begin{align}
\label{linklim}
\lim_{\rho \rightarrow \infty} {\sqrt{\frac{\rho}{6}}\av{\bL_0(x,x')}}&= 2\,\theta(x_0-x_0') \theta(\tau^2)
\delta(\tau^2)\\
& =  \theta(x_0-x_0') \delta(\tau^2)\\
& = 2 \pi G^{(4)}_0(x,x')\,.
\end{align}
This therefore suggests that we take the massless Green function on a flat 4-d causal set to be 
\begin{equation}
\label{massless4d}
K^{(4)}_0(x,x')= \frac{1}{2 \pi} \sqrt{\frac{\rho}{6}} L_0(x,x')\,.
\end{equation}
The relationship with the continuum Green function is not so direct as
in $d=2$ since here it is only in the continuum limit as $\rho\rightarrow \infty$ that the
mean of $K^{(4)}_0$ over sprinklings equals the continuum $G^{(4)}_0$.
We use this $K^{(4)}_0$  to construct a massive Green function $K^{(4)}_m(x,x')$ via
(\ref{convK}) as before.

Let a $k$-\textit{path between} $x'$ \textit{and} $x$ in a causal set  $C$  be  a $k$-chain between
$x'$ and $x$, $x'\prec x_1\prec
x_2\prec....x_{k-1}\prec x_k\prec x$,  in which all these
relations are links. For $k \ge 1$, define $L_k(x,x')$ to be the number of $k$-paths between $x'$ and $x$
when $x'\prec x$ and zero when $x'\not\prec x$.
The $L_k$'s are powers of the link matrix:
\begin{equation}
\label{numberlinks}
L_k(x,x') = \underbrace{L_0 * L_0* \ldots L_0}_{k+1} (x,x')\,.
\end{equation}
This gives
\begin{equation}
\label{csdiscreteGF4d}
K^{(4)}_m(x,x') =  \sum\limits_{k=0}^\infty  \biggl(-\frac{m^2}{\rho}\biggr)^{k}\biggl(\frac{1}{2\pi}\sqrt{\frac{\rho}{6}}\biggr)^{k+1} L_k(x,x')\,,
\end{equation}
where the sum terminates for each pair $x$ and $x'$.

For each two points $x$ and $x'$ of $\mink^2$  and each $k$, the random variable
$\bL_k(x,x')$ is $L_k(x,x')$ evaluated on a sprinkled causal set including $x$ and $x'$,
and hence we have the random variable $\bK^{(4)}_m(x,x')$:
 \begin{equation}
\label{discreteGF4d}
\bK^{(4)}_m(x,x') \equiv   \sum\limits_{k=0}^\infty  \biggl(-\frac{m^2}{\rho}\biggr)^{k}\biggl(\frac{1}{2\pi}\sqrt{\frac{\rho}{6}}\biggr)^{k+1} \bL_k(x,x').
\end{equation}
The limit as $\rho\rightarrow \infty$ of its mean  is equal to the series for the continuum Green function since
\begin{equation}
\label{abundancepaths}
 \av{\bL_k(x,x')} = \rho^k (\underbrace{\av{\bL_0}\ast \ldots \ast \av{\bL_0}}_{k+1}) (x,x')
\end{equation}
and so
\begin{align}
\lim_{\rho \rightarrow \infty} \av{\bK^{(4)}_m(x,x')} & = \lim_{\rho \rightarrow \infty} \sum\limits_{k=0}^{\infty} \biggl(-\frac{m^2}{\rho}\biggr)^{k}\biggl(\frac{1}{2\pi}\sqrt{\frac{\rho}{6}}\biggr)^{k+1} \av{\bL_k(x,x')}\\
&= \lim_{\rho \rightarrow \infty}  \sum\limits_{k=0}^{\infty} (-m^2)^{k}\biggl(\frac{1}{2\pi}\sqrt{\frac{\rho}{6}}\biggr)^{k+1}
\underbrace{\av{\bL_0}\ast \ldots \ast \av{\bL_0}}_{k+1} (x,x')\\
& = \sum\limits_{k=0}^{\infty} (-m^2)^{k}\underbrace{G^{(4)}_0 * G^{(4)}_0* \ldots G^{(4)}_0}_{k+1} (x,x')\\
&= G^{(4)}_m(x,x')\,.
\end{align}

Johnston interpreted (\ref{massless4d}) as a sum over paths between $x$ and $x'$. The hop-stop weights can be read off from Eqn (\ref{discreteGF4d}) as
$a=\frac{1}{2 \pi} \sqrt{\frac{\rho}{6}} $ and $b=-\frac{m^2}{\rho}$, respectively.

\begin{figure}[h]
 \centering
  \includegraphics[scale=0.5]{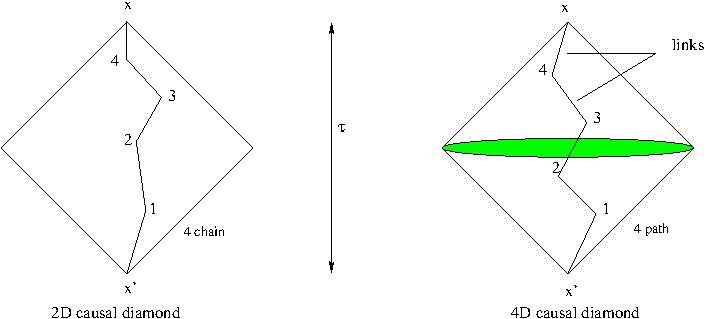}
\caption{The causal trajectories in $d=2$ and $4$ dimensions.}
 \end{figure}

\section{Generalisations}
\label{sec:generalisations}
The key to the above construction of a massive Green function is knowing the massless one. We can repeat it
if we can find the massless retarded Green function for causal sets sprinkled into more general curved spacetimes.

Consider the more general scalar theory with nonminimal coupling with Green function
$G_{m,\xi}(x,x')$ which satisfies
\begin{equation}
\label{curvedG}
(\Box_g - m^2 -\xi R)G_{m,\xi}(x,x')=\frac{1}{\sqrt{-g(x)} }\delta(x-x')\,.
\end{equation}
$G_{m,\xi}(x,x')$ can be obtained from $G_{0,\xi}(x,x')$ using the same series expansion Eqn(\ref{conv}):
\begin{equation}
\label{convxi}
G_{m,\xi}= \sum_{k=0}^\infty(-m^2)^k \underbrace{G_{0,\xi} * G_{0,\xi}* \ldots G_{0,\xi}}_{k+1} \,.
\end{equation}

In the special case when the spacetime has constant scalar curvature $R$, then the $\xi R$ term just modifies the mass and  $G_{m,\xi}(x,x')$  can be obtained from
the minimally coupled massless Green  function $G_{0,0}(x,x')$
using a series expansion Eqn(\ref{conv}) with $m^2$ replaced by $m^2+\xi R$.
In general, for constant $R$, we can relate the two Green functions 
\begin{equation}
\label{cconvxi}
G_{m', \xi'}= \sum_{k=0}^\infty(-{m'}^2 - \xi' R +m^2 + \xi R)^k \underbrace{G_{m,\xi} * G_{m,\xi}*
  \ldots G_{m,\xi}}_{k+1} \, , 
\end{equation}
for any $(m,\xi)$, $(m',\xi')$. 

We seek analogous massive scalar Green functions, $K_{m,\xi}(x,x')$, for causal sets sprinkled into
nonflat spacetimes. We will see that this is possible in special cases.

\subsection{$d=2$}

Every $d=2$  spacetime is locally conformally flat. The conformal coupling in $d=2$ is $\xi=0$, \textit{i.e.}
conformal coupling is minimal coupling. If the spacetime is topologically trivial and
consists of one patch covered by conformally flat
coordinates, then the minimally coupled massless Green function equals the flat spacetime Green function (\ref{masslesscont2d}).

 Therefore, we propose that on
  causal sets sprinkled into such $d=2$ spacetimes,
   the massless minimally coupled causal set Green function,
  $K^{(2)}_{0,0}(x,x')$, is the flat one given by Eqn (\ref{massless2d}) and therefore that
  $K^{(2)}_{m,0}(x,x')$ is the flat one
given by Eqn (\ref{discreteGF2d}):
\begin{equation}
K^{(2)}_{m,0}(x,x') =  \sum\limits_{k=0}^\infty  \biggl(- \frac{m^2}{\rho}\biggr)^{k}\biggl(\frac{1}{2}\biggr)^{k+1} C_k(x,x')\,,
\end{equation}

The argument that the mean over sprinklings of the corresponding random variable will be the
correct continuum Green function proceeds exactly as in the flat case: (\ref{discreteGF2d})--(\ref{proof3}).
However it is formal and we will provide more concrete evidence. We will verify
directly that this $K^{(2)}_{m,0}$ does have the correct mean value over sprinklings in a
Riemann Normal Neighbourhood (RNN).

In our calculation below as well as in Section \ref{subsec:4d},   the RNN should be seen as providing an
intermediate scale at which the continuum description is still valid, 
and which is therefore  much larger than the discreteness scale. The
reason to use the RNN is simply that the calculations can be done
explictly to leading order both in the causal set as well as the
continuum.

\subsubsection{ RNN in $d=2$}

Consider the RNN $(O,g)$ with Riemann normal coordinates with origin $x'$.
The metric at $x\in O$ can be expanded to first order about $x'$ in these coordinates as
\begin{equation}
\label{rnn}
g_{ab}(x)=\eta_{ab}+\dfrac{1}{2!}\partial_{c}\partial_{d}\,g_{ab}(x')x^{c}x^{d}+\mathcal{O}(x^{3}).
\end{equation}
where $\eta_{ab}$ is the metric of Minkowski spacetime in inertial coordinates and
$\partial_c g_{ab}(x')=0$. We assume that
the RNN is approximately flat, $\textit{i.e.}$
$|R \tau^2(x,x') |<<1$, and work in an approximation where we drop terms involving derivatives of
the curvature or quadratic and higher powers of the curvature.

The $d$ dimensional momentum space Green function in a RNN has been calculated by Bunch and
Parker~\cite{bunchparker}. To leading order the density
\begin{equation}
\bG_{m,\xi}(x,x')\equiv (-g(x))^{\frac{1}{4}}G_{m,\xi}(x,x')
\end{equation}
satisfies the  equation
\begin{equation}
\label{rnnbox}
(\Box_\eta -(m^2+(\xi-\frac{1}{6}) R(x'))\bG_{m,\xi}(x,x')\approx - \delta(x-x')\,,
\end{equation}
where $\Box_\eta = \eta^{ab}\nabla_a\nabla_b$ and acts on the $x$ argument. 
This
has the momentum space solution
\begin{equation}
\label{ftG}
\bG_{m,\xi}(p) \approx \frac{1}{p^2+m^2} - (\xi -\frac{1}{6}) R(x')\frac{1}{(p^2+m^2)^2}.
\end{equation}
This solution was obtained iteratively using the expansion 
\begin{equation} 
\bG_{m,\xi}(p)=\bG_{m,\xi,0}(p)+\bG_{m,\xi,1}(p)+\bG_{m,\xi,2}(p) + \ldots 
\end{equation} 
where $\bG_{m,\xi,0}(p)=(k^2+m^2)^{-1}$ is the flat spacetime Green function which is independent of
$\xi$. This expansion is valid when the Compton wavelength of the particle is much smaller than the
curvature scale, i.e., $m^2>> \xi R$, a physically reasonable assumption. 
The spacetime function  can then be expressed as
\begin{equation}
\bG_{m,\xi}(x,x') \approx G^F_m(x,x')+\frac{1}{2m} (\xi -\frac{1}{6}) R(x')\, \partial_m G^F_m(x,x'),
\end{equation}
where $G_m^F(x,x')$ is the massive minimally coupled Green function in $\mink^d$.
The Green  function is then
\begin{equation}
\label{rnnG}
G_{m,\xi}(x,x')\approx \biggl(1+\frac{1}{12} R_{ab}(x')x^a x^b\biggr)G^F_m(x,x')+ \frac{1}{2m} (\xi -\frac{1}{6}) R(x')\, \partial_m G^F_m(x,x').
\end{equation}

Now we specialise to $d=2$. Using the $d=2$ Minkowski spacetime solution for the massive retarded solution
\begin{equation}
 \frac{1}{2} \theta(x_0) \theta(\tau^2) J_0(m\tau)\,,
\end{equation}
where $\tau=\tau(x,x')$ (\ref{tau2d}), the retarded massive Green  function
in $(O,g)$ is given by 
\begin{equation}
\label{2dcontden}
G^{(2)}_{m,\xi}(x,x') \approx \theta(x_0) \theta(\tau^2)\left[  \dfrac{1}{2}J_{0}(m\tau) +\dfrac{R(x')\tau^2}{48}J_{2}(m\tau)
- \frac{\xi R(x')\tau}{4m}J_1(m\tau)\right]\,.
\end{equation}

Let us define
\begin{equation}
\label{2dgeneralab}
{\cal{K}}^{(2)}(a,b)(x,x') \equiv \sum\limits_{k=0}^\infty  a^{k+1} b^k C_k(x,x')
\end{equation}
for arbitrary weights $a$ and $b$. We want to show that the corresponding random variable for sprinklings into a RNN has the correct mean value, (\ref{2dcontden}) when $a$ and $b$ take their
flat space values $a= \frac{1}{2}$ and $b = -\frac{m^2}{\rho}$.

We can calculate $\av{{\cal{K}}^{(2)}(a,b)(x,x')}$ starting from  Eqn (\ref{2dgeneralab}) if we know
 $\av{\bC_k(x,x')}$ in a small causal diamond. This was
calculated, to first order in curvature, in \cite{rss} for arbitrary $d\geq 2$.  In $d=2$ the expression is
\begin{equation}
\av{\bC_{k}(x,x')}\approx \av{\bC_{k}(x,x')}_\eta\bigg(1-\dfrac{R(x')\tau^2}{24}\dfrac{k}{k+1}\bigg)
\end{equation}
where $\av{\bC_{k}(x,x')}_\eta= \theta(x_0)\theta(\tau^2)\dfrac{1}{\Gamma(k+1)^2}\biggl(\dfrac{\rho\tau^2}{2}\biggr)^k$
is the mean in flat space.
Using the series expansion of the Bessel functions we see that
\begin{eqnarray}
\label{2csprop}
\av{{\cal{K}}^{(2)}(a,b)} &\approx& \theta(x_0)\theta(\tau^2) \sum_{k=0}^{\infty}a^{k+1}b^{k}\bigg(\dfrac{\rho \tau^2}{2}\bigg)^{k}\dfrac{1}{(\Gamma(k+1))^2}\bigg(1-\dfrac{R(x')\tau^2}{24}\dfrac{k}{k+1}\bigg)\nonumber\\
&\approx& \theta(x_0)\theta(\tau^2) \left[aI_{0}(\tau\sqrt{2ab\rho}) - \dfrac{aR(x')\tau^2}{24}I_{2}(\tau\sqrt{2ab\rho})\right].
\end{eqnarray}
If we set
$a= \frac{1}{2}$,  $b= - \frac{m^2}{\rho}$ we find
\begin{equation}
\label{2csprop1}
\av{{\cal{K}}^{(2)}(\frac{1}{2}, - \frac{m^2}{\rho})(x,x')}\approx  \theta(x_0)\theta(\tau^2)\left[\dfrac{1}{2}J_{0}(m\tau) +\dfrac{R(x')\tau^2}{48}J_{2}(m\tau)\right],
\end{equation}
which matches  Eqn (\ref{2dcontden}) for $\xi=0$.

We further note that in the RNN since $R(x') \approx R$,  a constant  to this order of
approximation,  we can use the
observation above that $\xi R$ can be treated as a contribution to the mass.
Putting $a=\frac{1}{2}$ and  $b=-\frac{(m^2+\xi R)}{\rho}$ in (\ref{2csprop})  and using $m^2 >>
\xi R$, we obtain
\begin{align}
& \theta(x_0)\theta(\tau^2)\left[\dfrac{1}{2}J_{0}(\tau\sqrt{m^2+\xi R})+ \dfrac{R\tau^2}{48}J_{2}(\tau\sqrt{m^2+\xi R})\right]\nonumber\\
\approx &\  \theta(x_0)\theta(\tau^2)\left[\frac{1}{2}\sum_{n=0}^{\infty}\frac{(-1)^n}{(n!)^2}\bigg(\frac{\tau}{2}\bigg)^{2n}(m^2+\xi R)^n + \dfrac{R\tau^2}{48}J_{2}(m\tau)\right]\nonumber\\
\approx&\  \theta(x_0)\theta(\tau^2)\left[\dfrac{1}{2}J_{0}(m\tau)  + \dfrac{R\tau^2}{48}J_{2}(m\tau)
 - \frac{\xi R\tau}{4m}J_1(m\tau)\right]
\end{align}
which agrees with Eqn (\ref{2dcontden}).   Thus, for a causal set sprinkled into an approximately flat causal diamond in $d=2$,
Eqn (\ref{2dgeneralab}) with $a=  \frac{1}{2}$ and $b= - \frac{(m^2 + \xi R(x'))}{\rho}$  is approximately the
``right'' massive causal set Green function for general coupling $\xi$.

\subsection{$d=4$}

\subsubsection{RNN  in $d=4$ }
\label{subsec:4d}

The approximate continuum retarded Green  function in the RNN in $d=4$ simplifies to
\begin{align} \label{rnn4dcont}
G^{(4)}_{m,\xi}(x,x') \approx &\ \theta(x_0)\Biggl[  \biggl(\dfrac{1}{2\pi}\delta(\tau^2) -
\theta(\tau^2)\frac{m}{4\pi\tau}J_1(m\tau)\biggr)\biggl (1+
\dfrac{1}{12}R_{ab}(x')x^a x^b\biggr) \Biggr]\\
& - \theta(x_0)\theta(\tau^2)\bigg(\xi-\frac{1}{6}\bigg)\frac{R(x')}{8\pi}J_0(m\tau),
\end{align}
which reduces to the massless Green  function
\begin{align}
\label{rnn4dcont0}
 G^{(4)}_{0,\xi}(x,x') \approx
 \dfrac{1}{2\pi} \theta(x_0)\delta(\tau^2)\bigg(1+\dfrac{1}{12}R_{ab}(x')x^a x^b\bigg) - \theta(x_0)\theta(\tau^2)\bigg(\xi-\frac{1}{6}\bigg)\frac{R(x')}{8\pi}.
\end{align}
Even this simplified expression is formidable to mimic in the causal set since not only does it
require the discrete scalar curvature \cite{bd} but also the \textit{components} of the Ricci curvature for
which no expression is known.  However,  for conformal coupling
$\xi=\frac{1}{6}$ and Einstein spaces with Ricci curvature $R_{ab}\propto g_{ab}$,
(\ref{rnn4dcont0}) reduces to the Minkowski spacetime form (\ref{cont4d}). Indeed, we only require
that $R_{ab}(x')\appropto g_{ab}(x') $ upto the order we are considering. 
This suggests that the flat spacetime massless causal set Green function (\ref{massless4d})
may give the right continuum Green function.  Since $R$ is approximately constant
in the RNN (and exactly constant in an Einstein space)  we can
use the series in powers of the massless Green function to propose the
massive one for arbitrary $\xi$.

For the massless field let us calculate the mean of the link matrix, given by (\ref{linkexp}).  The
spacetime volume in the RNN has corrections to the Minkowski spacetime volume $V_\eta(x,x')$
\cite{myrheim, gs,ks} which in $d=4$ are
\begin{equation}
V(x,x')\approx V_\eta(x,x')\bigg(1-\dfrac{1}{180}R(x')\tau^2+\dfrac{1}{30}R_{ab}(x')x^ax^b\biggr).
\end{equation}
To leading order then
\begin{eqnarray}
\av{\bL_0(x,x')} \approx \theta(x_0) \theta( \tau^2) e^{-\rho V_\eta(x,x')}\bigg(1+\dfrac{\rho V_\eta(x,x')}{180}R(x')\tau^2-\dfrac{\rho V_\eta(x,x')}{30}R_{ab}(x')x^ax^b\bigg).
\end{eqnarray}
Since  $V_\eta(x,x')=\frac{\pi}{24}\tau^4(x,x')$,  $\sqrt{\rho} \av{\bL_0(x,x')}$ contains
terms of the form
\begin{equation}
h_n(\rho,\tau) \equiv \sqrt{\rho} (c \rho \tau^4)^n \exp(-c\rho\tau^4),
\end{equation}  with $n=0,1$.
As shown in Appendix A
\begin{equation}
\lim_{\rho\to\infty}h_n(\rho,\tau)=\dfrac{\Gamma(n+1/2)}{2\sqrt{c}}\delta(\tau^2).
\end{equation}
Using this,  we find that
\begin{equation}
\label{rnncs4d}
\lim_{\rho \rightarrow \infty} \frac{\sqrt{\rho}}{2\pi \sqrt{6}} \av{\bL_0(x,x')} \approx
\dfrac{1}{2\pi}\theta({x_0})\delta(\tau^2)\bigg(1+\dfrac{R(x')\tau^2}{360}-\dfrac{R_{ab}(x')x^ax^b}{60}\bigg).
\end{equation}
The second term vanishes in general and so does the third term when $R_{ab}(x') \propto g_{ab}(x')$
upto this order, and we recover (\ref{cont4d}).  Thus, for sprinklings into a RNN with $R_{ab}(x')
\propto g_{ab}(x')$ to this order, the continuum limit of the mean of (\ref{massless4d}) is
approximately the correct value for the Green function of the conformally coupled massless field.

Defining 
\begin{equation}
\label{4dgeneralab}
{\cal{K}}^{(4)}(a,b)(x,x') \equiv \sum\limits_{k=0}^\infty  a^{k+1} b^k L_k(x,x')\,,
\end{equation}
we propose that this is the appropriate  causal set Green function for  the massive field and
arbitrary coupling $\xi$  in an  RNN with $R_{ab}(x') \propto g_{ab}(x')$ to this order,   for $a=
\frac{1}{2\pi}\sqrt{\frac{\rho}{6}}$ and $b = -\frac{m^2+ (\xi -\frac{1}{6})R}{\rho}$.

We are unable to verify this directly as we lack knowledge about the mean of
$\bL_k$, the number of $k$-paths for $k\ge 1$, even in a RNN.

\subsubsection{$d=4$ de Sitter and anti de Sitter}
\label{subsec:desitter}

In $d=4$ for  conformally flat spacetimes $g_{ab}=\Omega^2(x)\eta_{ab}$   the conformally
coupled massless Green  function is related to that in $\mink^4$ by  
\begin{equation} 
\label{confflat}
G_{0,\xi_c}(x,x')=\Omega^{-1}(x) G_0^{F}(x,x')\Omega^{-1}(x'),  
\end{equation} 
where $\xi_c = \frac{1}{6}$ and  $G_0^F(x,x')$ denotes the retarded massless Green  function in $\mink^4$. 
When $g_{ab}$ in
addition has constant scalar curvature the massive Green function for arbitrary $\xi$ can be obtained from
$G_{0,\xi_c}(x,x')$  using Eqn (\ref{convxi}).

An example is the conformally flat patch of de Sitter spacetime 
\begin{equation} 
ds^2= \frac{1}{(1+H x_0)^2} \biggl( -d x_0^2 + \sum_{i=1}^3 d x_i^2  \biggr), 
\end{equation} 
where $x_0 $ is the conformal time ($-\frac{1}{H}< x_0 <\infty $) and $H=\sqrt{\frac{\Lambda}{3}}$
with $\Lambda$ the cosmological constant.  The conformally coupled massless retarded Green  function is 
\begin{equation} 
G_{0,\xi_c}(x,x')=\frac{1}{2\pi} \theta(t-t')\delta(\tau^2(x,x')) (1+Hx_0)(1+Hx_0'). 
\end{equation} 
Since de Sitter is homogeneous, one can choose $x'$ to lie at the  convenient location  $x'=(0, \vec 0)$ so that 
\begin{equation}
 G_{0,\xi_c}(x,x')=\frac{1}{2\pi} \theta(x_0)\delta(\tau^2(x,x')) (1+Hx_0). 
\end{equation} 

Taking our cue from the RNN calculation, we look to the link matrix $L_0(x,x')$ whose expectation 
value is given by Eqn (\ref{linkexp}). Taking $x'=(0,\vec 0)$, in the limit 
\begin{equation} 
\label{sqrtv}
\lim_{\rho \rightarrow \infty} \sqrt{\frac{\rho}{\pi}} \av{\bL_0(x,x')}=\theta(x_0)
\theta(\tau^2(x,x'))\delta(\sqrt{V(x,x')}).  
\end{equation} 
In order to evaluate this expression we need to find $V(x,x')$.  In \cite{gs} this volume
was calculated for a large interval when $\vec x=\vec x'$. However,  it is the
small volume limit that is relevant to our present  calculation.  When $x$  lies
in an RNN about $x'$, the calculation in the previous section suffices. However, we also need to consider
intervals of small volume that lie outside of the RNN. These ``long-skinny'' intervals hug the future
light cone of $x'$ and it is this contribution to  Eqn (\ref{sqrtv}) that we will now consider. 

In the following light cone coordinates 
\begin{equation} 
u=\frac{1}{2}(x_0-x_3), v=\frac{1}{2}(x_0+x_3), 
\end{equation} 
let $u(x)=\epsilon$, $v(x)=L$, with $\alpha^2\equiv \frac{\epsilon}{L} <<1$.  Since there is a
spatial rotational symmetry in de Sitter, we can also take $x_1=x_2=0$.  In order to simplify the
calculation of $V(x,x')$, we perform a boost about $x'$ in the $x_0-x_3$ 
plane about $x'$ so that  $\tx=(\tx_0, \vec{0})$. The boost parameter  is then $\beta=\frac{x_3}{x_0} \approx
1-2\alpha$.  In these coordinates the conformal factor at a point $y=(y_0, \vec{y})$ is  
\begin{equation} 
\tO^2(\ty)\approx \frac{1}{(1+A(\ty_0+\ty_3))^2}
\end{equation} 
where $A=\frac{1}{2}H\alpha$. Further transforming to cylindrical coordinates $(\ty_1,\ty_2,\ty_3)
\rightarrow (r,\phi, \ty_3)$ we can split $V(x,x')$ into two multiple integrals
\begin{eqnarray} 
\label{volexprs} 
V_I(x,x')&=& \int_{0}^{-\frac{\tau}{2}} \!\!d\ty_0 \int_{-\ty_0}^{\ty_0} \!\!\!\! d\ty_3\int_0^{\sqrt{\ty_0^2-\ty_3^2}}
\!\! \!\!\!r dr \int_0^{2\pi} d \phi \, \biggl(1+A(\ty_0+\ty_3)\biggr)^{-4}   \\ 
V_{II}(x,x')&=& \int_{\frac{\tau}{2}}^{\tau} \!\! d\ty_0 \int_{-\tau+\ty_0}^{\tau-\ty_0} \!\!\!\! d\ty_3\int_0^{\sqrt{(\tau-\ty_0)^2-\ty_3^2}}
\!\! \!\!\!\!\! \!\!\!r dr \int_0^{2\pi} d \phi \,\, \biggl(1+A(\ty_0+\ty_3)\biggr)^{-4}    
\end{eqnarray} 
with $V(x,x')=V_I(x,x')+V_{II}(x,x')$. Evaluating these expressions using $\tau^2=4 L \epsilon$ we find that 
\begin{equation}
\label{voldS}
\sqrt{V(x,x')} = \frac{1}{2}\sqrt{\frac{\pi}{6}}\frac{\tau^2}{{(1+A\tau)}}\approx
\frac{1}{2}\sqrt{\frac{\pi}{6}}\biggl(\frac{4 L \epsilon}{1+HL} \biggr), 
\end{equation} 
which substituted into Eqn
(\ref{sqrtv}) gives 
\begin{equation} 
\lim_{\rho \rightarrow \infty} \sqrt{\frac{\rho}{\pi}} \av{\bL_0(x,x')}=\theta(t-t')
\theta(\tau^2(x,x')) \sqrt{\frac{6}{\pi}}(1+HL)  \delta(4 L \epsilon). 
\end{equation} 
In the small $\alpha$ limit the conformally coupled de Sitter Green  function is 
\begin{equation}
  G_{0,\xi_c}(x,x') \approx \frac{1}{2\pi} (1+H L) \theta(x_0)\delta(4L\epsilon)  
\end{equation} 
and hence 
\begin{equation}
\label{masslessdeS}
 \lim_{\rho \rightarrow \infty} \frac{1}{2\pi}\sqrt{\frac{\rho}{6}}
 \av{\bL_0(x,x')}=G_{0,\xi_c}(x,x'). 
\end{equation} 

As in the RNN, defining 
\begin{equation}
\label{deS}
{\cal{K}}^{(4)}(a,b)(x,x') \equiv \sum\limits_{k=0}^\infty  a^{k+1} b^k L_k(x,x')\,,
\end{equation}
we propose that this is the appropriate  causal set Green function for  the massive field and
arbitrary coupling $\xi$  in de Sitter spacetime for  $a=
\frac{1}{2\pi}\sqrt{\frac{\rho}{6}}$ and $b = -\frac{m^2+ (\xi -\frac{1}{6})R}{\rho}$. 

Although our calculation is restricted to the conformally flat patch of de Sitter spacetime, the
result applies to global de Sitter, for the following reason. Let $x' \prec x$ in (global) de Sitter
spacetime. Consider a Lorentz transformation about $x'$ in the 5-dimensional Minkowski spacetime in
which the hyperboloid that is de Sitter spacetime is embedded, which brings $\vec x=\vec x'$.  This
transformation preserves the hyperboloid. One can then choose the conformally flat patch of de
Sitter with origin $\vec x'$, and use the above construction. When $x,x'$ are not causally related,
the Green functions vanish in both cases.  Thus the Green function for global de Sitter is retarded
if the conformally flat Green function is, and both satisfy the same equations, because there is no
“wrap-around” in de Sitter.

The causal set Green function  we propose is well defined on a sprinkling into global de
Sitter.  Moreover, as we have shown its continuum limit matches that of the Green
function into the conformally flat patch and thence from the above discussion, also the  Green function
of global  de Sitter spacetime.

In anti de Sitter (adS) spacetime there exist pairs of events $x'\prec x$ such that $\tau(x,x') $ is
finite, but $V(x,x')$ is infinite. While it is possible to obtain a Poisson sprinkling into such a
spacetime, the resulting poset is not locally finite and hence not strictly a causal set. Such an
interval is moreover not globally hyperbolic and hence falls outside the scope of our analysis.
However, the interior of a conformally flat patch of adS (the so-called half-space) is globally
hyperbolic and moreover, $V(x,x')$ is finite for every $x'\prec x$ in this region. Hence this patch
of adS has a causal set description.

In the conformally flat patch the adS metric takes the form 
\begin{equation} 
ds^2= \frac{1}{(1+H x_3)^2} \biggl( -d x_0^2 + \sum_{i=1}^3 d x_i^2  \biggr), 
\end{equation} 
where we have off set the coordinates $x_3 \rightarrow x_3 + \frac{1}{H}$ in order to connect  with the
de Sitter calculation. Again choosing $x'=(0, \vec 0)$, we can write the massless Green function as 
\begin{equation}
 G_{0,\xi_c}(x,x')=\frac{1}{2\pi} \theta(x_0)\delta(\tau^2(x,x')) (1+Hx_3). 
\end{equation} 
In the boosted coordinates, upto order $\alpha^2$, the conformal factor
\begin{equation} 
\Omega^2(y)=\frac{1}{(1+Hy_3)^2} \approx \frac{1}{(1+A(\ty_0+\ty_3))^2}
\end{equation} 
and is identical to that of de Sitter in the  calculation above. Moreover, to this order,
$(1+Hx_3)=(1+H(L-\epsilon)) \approx (1+HL)$, so that $\sqrt{V(x,x')}$
is given by Eqn (\ref{voldS}). The same argument can then be carried through to show that the massive causal set
Green function for arbitrary $\xi$ in the conformally flat patch of de Sitter is given by Eqn (\ref{deS}). 

We have thus proved \textit{exact}  results in de Sitter spacetime and in a conformally flat patch of anti de Sitter spacetime, namely that
the mean of the causal set retarded Green function 
\begin{equation} 
\bK_0(x,x')=\frac{1}{2\pi}\sqrt{\frac{\rho}{6}}
 \av{\bL_0(x,x')}
\end{equation} 
is equal to the continuum massless conformally coupled Green function in the limit $\rho\rightarrow
\infty$.  In addition, we make the proposal that the limit of the mean of 
${\cal{K}}^{(4)}(a,b)(x,x') $ with the appropriate $a$ and $b$ is the continuum massive Green
function for arbitrary conformal coupling $\xi$.

\section{Discussion} 
\label{sec:discussion} 

We have demonstrated that Johnston's hop-stop model for a Green function on a causal set can be
generalised whenever an  identification can be made of  the massless retarded Green function.

As a final  illustration, we make a proposal for the causal set Green  function
in $d=3$ Minkowski spacetime.   In continuum flat spacetime in 3 dimensions the massless scalar propagator is
\begin{equation} 
\label{3dgf}
G^{(3)}_0(x,x')=\theta(t-t')\theta(\tau^2)\dfrac{1}{2\pi\tau(x,x')}, 
\end{equation} 
where for now we ignore the singular behaviour at $\tau(x,x')=0$.  The causal set counterpart of the
proper time $\tau(x,x')$ in ${}^d\mink$ was given by Brightwell and Gregory \cite{bg} to be
proportional to the \textit{length}  $l(x,x')$ of the longest chain (LLC) from $x'$ to $x$, where 
{length} of a $k$-chain is taken to be $k+1$.  Explicitly
\begin{equation}
\label{bg1}  
\lim_{\rho \rightarrow \infty} \av{\bl(x,x')} (\rho V(x,x'))^{-1/d} = m_d
\end{equation} 
where $m_d$ is a dimension dependent constant bounded by
\begin{equation} 
\label{bounds}
1.77\leq\frac{2^{1-\frac{1}{d}}}{\Gamma(1+\frac{1}{d})}\leq
m_d\leq\frac{2^{1-\frac{1}{d}}e\,(\Gamma(1+d))^{\frac{1}{d}}}{d}\leq 2.62.
\end{equation} 
In ${}^d\mink$, $\rho V(x,x')= \zeta_d \tau^d(x,x')$ with $\zeta_d$ a dimension dependent constant, so that 
\begin{equation}
\label{bg}
\lim_{\rho \rightarrow \infty} \rho^{-\frac{1}{d}} \av{\bl(x,x')}=\kappa_d\,\tau(x,x')
\end{equation}
where $\kappa_d\equiv m_d(\zeta_d)^{1/d}$.  This \textit{suggests} that  the $d=3$ massless Green  function on $C$ is 
\begin{equation} 
\label{3dcsprop}
K_0(x,x') \equiv a H_0(x,x'). 
\end{equation} 
where 
\begin{equation}
\label{def}
H_0(x,x')\equiv
\left\{
	\begin{array}{ll}
		\dfrac{1}{l(x,x')}  & \mbox{if } x' \prec x\\
		0 & \mbox{} \mathrm{otherwise}.
	\end{array}
\right.
\end{equation}
This will give us the desired $d=3$ Green function if it were also true that    
\begin{equation}
\label{limit3d}
\lim_{\rho \rightarrow \infty} \bigg\langle\frac{1}{l(x,x')}\bigg\rangle \rho^{\frac{1}{3}}= \frac{1}{\kappa_3\tau(x,x')}
\end{equation}
then comparison with Eqn (\ref{3dgf}) gives $a=\rho^{1/3} \frac{\kappa_3}{2\pi}=(\frac{\rho \pi}{12})^{1/3} \frac{m_3}{2\pi}$. \\ 
While we do not have an analytical proof of Eqn (\ref{limit3d}), simulations shown in Appendix B demonstrate that for large $\rho$, $ \av{\frac{1}{l(x,x')}}\rightarrow\frac{1}{\av{l(x,x')}}$ and hence Eqn (\ref{limit3d}) is indeed a good approximation. 

In order to extend this to the massive case, we need to ask what the analogue of the convolution in
(\ref{numberchains}) is. Because of the non-trivial weight $\frac{1}{l(x,x')}$, we cannot simply
count chains to get $C_k$. The convolution 
\begin{equation} 
\av{\bHH_0}*\av{\bHH_0}=\rho \int d^3x_1 \av{\bHH_0(x,x_1)}{\av{\bHH_0(x_1,x')}}=\av{\bHH_1(x,x')}
\end{equation} 
where 
\begin{equation} 
H_1(x,x')=\sum_{x_1}H_0(x,x_1)H_0(x_1,x')=\sum_{x_1}\dfrac{1}{l(x,x_1)}C_0(x,x_1)\dfrac{1}{l(x_1,x')}C_0(x_1,x'). 
\end{equation} 
counts instead the number of $1$-chains weighted by the inverse of the length of the longest
possible chain in $C$ between each successive pair of joints in the given chain. 
As in $d=2$ the trajectories are chains, but the $H_k(x,x')$ are not
obtained by merely counting chains; each $k$-chain is weighted by the inverse of the length of the
longest possible chain $C$ between each pair of joints in the given chain. 

Again defining 
\begin{equation}
\label{3dgeneralab}
{\cal{K}}^{(3)}(a,b)(x,x') \equiv \sum\limits_{k=0}^\infty  a^{k+1} b^k H_k(x,x')\,,
\end{equation}
we propose that this is the appropriate  causal set Green function for  the massive field for  $a=(\frac{\rho \pi}{12})^{1/3} \frac{m_3}{2\pi}$
and $b = -\frac{m^2}{\rho}$. 

In \cite{johnstonthesis} a proposal for the d=3 Green function was made using the relationship
between $\tau(x,x')$ and the volume $V(x,x')$, $\tau(x,x')\propto V(x,x')^{\frac{1}{3}}$. Our
proposal uses instead the causal set analogue of $\tau(x,x')$ directly. In the large $\rho$ limit,
one would expect both proposals to give the same result. 

In principle, the causal set Green functions in $d>4$ can also be obtained. Massless propagators in the continuum are derivatives of either $\frac{1}{\tau}$ or
$\delta(\tau^2)$ depending on whether $d$ is odd or even. Since derivatives of $\delta(\tau^2)$ of
any order can always be written as products of $\delta(\tau^2)$ and $\frac{1}{\tau}$, the knowledge
of the causal set analogues of these two quantities with appropriate weights suffices  to write
down the discrete propagator.  We leave this to future work. 

Another aspect we have ignored in this work are the causal set induced corrections to the retarded
Green  functions. Given that causal set theory posits a fundamental discreteness, the $\rho
\rightarrow \infty$ limit is only a mathematical convenience. Indeed it is the large $\rho$
corrections to the continuum Green function which are phenomenologically interesting. This has been
explored in \cite{Johnstoncorrectionterms} for cases of  $d=2$ and $4$ Minkowski spacetime.  Also, while
our analysis has focused on the mean of the causal set Green function, we have not analysed the
fluctuations.  We hope to be able to address these issues in the future.

\section{Acknowledgements} 

This work was  supported in part under an agreement with Theiss Research and funded by a
grant from the FQXI Fund on the basis of proposal FQXi-RFP3-1346 
to the Foundational Questions Institute. FD’s work is supported in part by STFC grant ST/L00044X/1. Simulations were performed using the \texttt{CausalSets} toolkit \cite{causet} within the \texttt{Cactus} high performance computing framework \cite{cactus}. NX would like to thank David Rideout and Simran Singh for helpful discussions on \texttt{Cactus}.   



\section{Appendix A} 
Define the function 
\begin{equation} 
h_n(\rho,z)\equiv \sqrt{\rho}(\pi c\rho z^2)^ne^{-\pi c\rho z^2}, 
\end{equation} 
{where $n\geq 0$}.  
We now show that 
\begin{equation}
\label{distid}
\lim_{\rho\to\infty}h_n(\rho, z)=\dfrac{\Gamma(n+1/2)}{\sqrt{\pi c}}\delta(z).
\end{equation}
First, we evaluate the integral
\begin{eqnarray} 
\int_{-\infty}^{\infty}dz\,\, h_n(\rho, z) &=& 
2\sqrt{\rho}\int_{0}^{\infty}dz\,\, (\pi c\rho z^2)^ne^{-\pi c\rho z^2} \nonumber \\ 
&=& \dfrac{1}{\sqrt{\pi c}}\int_{0}^{\infty}dt\,\, (t)^{n-1/2}e^{-t}\nonumber\\
&=&\dfrac{\Gamma(n+1/2)}{\sqrt{\pi c}}, 
\end{eqnarray} 
which is independent of $\rho$.  
Next, we integrate $h_n(\rho, z)$ with an analytic  test function and take the
limit $\rho\rightarrow \infty$.  If $f(z)$ is odd, the integral vanishes (this also happens with the
delta function) and we can restrict to even analytic functions 
\begin{equation} 
f(z)=\sum_{k=0}^{\infty}a_k z^{2k}.  \end{equation}  
For this, 
\begin{eqnarray}
\lim_{\rho\rightarrow\infty}\int_{-\infty}^{\infty}dz\,\,f(z)\,h_n(\rho, z)&=&\lim_{\rho\rightarrow\infty}\sum_{k=0}^{\infty}a_k\int_{-\infty}^{\infty}dz\,\,z^{2k}\,h_n(\rho, z)\nonumber\\
&=&\lim_{\rho\rightarrow\infty}2\sum_{k=0}^{\infty}a_k\int_{0}^{\infty}dz\,\,z^{2k}\sqrt{\rho}(\pi c\rho z^2)^ne^{-\pi c\rho z^2}\nonumber\\
&=&\lim_{\rho\rightarrow\infty}2\sum_{k=0}^{\infty}a_k\dfrac{\sqrt{\rho}}{(\pi c\rho)^k}\int_{0}^{\infty}dz\,\,(\pi c\rho z^2)^{n+k}e^{-\pi c\rho z^2}\nonumber\\
&=&\lim_{\rho\rightarrow\infty}\sum_{k=0}^{\infty}a_k\dfrac{\Gamma(n+k+1/2)}{\sqrt{\pi c}(\pi c\rho)^k}\nonumber\\
&=&a_0\dfrac{\Gamma(n+1/2)}{\sqrt{\pi c}}=\dfrac{\Gamma(n+1/2)}{\sqrt{\pi c}}f(0).\nonumber
\end{eqnarray}
Noting that  $n=0$ is   the  usual Gaussian integral, and that the behaviour with test functions is one way to define a delta function \cite{deltarefs}, this
proves Eqn (\ref{distid}).

\section{Appendix B}

Here we present simulations that provide  support for Eqn
(\ref{limit3d}) for large $\rho$. Starting with Eqn (\ref{limit3d})
in $d=3$ we see that 
\begin{equation} 
\lim_{N \rightarrow \infty} \bigg\langle\frac{1}{l(x,x')}\bigg\rangle
\bigg(\frac{N}{V(x,x')}\bigg)^{\frac{1}{3}}=
                                                                                                                     \frac{1}{m_3\,\zeta_3^{1/3}\tau(x,x')}
\end{equation} 
where we have used $\rho=\frac{N}{V}=\frac{N}{\zeta_3\,\tau^3}$ and
$\zeta_3=\frac{\pi}{12}$.  Since the volume $V(x,x')$ is fixed,  the
limit $\rho \rightarrow \infty$ is the same as $N \rightarrow \infty$
and hence this simplifies to 
\begin{equation} 
\lim_{N \rightarrow \infty} \bigg\langle\frac{1}{l(x,x')}\bigg\rangle=\frac{1}{m_3\,N^{1/3}}
\end{equation}
Using Eqn (\ref{bounds}) we see that 
\begin{equation}
b_l := \frac{1}{1.77\,N^{1/3}}\leq\lim_{N \rightarrow \infty}
\bigg\langle\frac{1}{l(x,x')}\bigg\rangle\leq\frac{1}{2.62\,N^{1/3}}
=: b_u
\end{equation}
where we have defined $b_l$ and $b_u$ as the lower and upper bounds
respectively.  

We calculate $\av{\frac{1}{\l(x,x')}}$ and  $\frac{1}{\av{\l(x,x')}}$
for sprinklings into a causal diamond in ${}^3\mink$, for  $N$ values
ranging from  $100$ to $50000$ in steps of $100$. For each $N$ value we perform  over 50
trials from which the averages are calculated. Our results are shown
in Figs (\ref{plot_1})-(\ref{plot_5}). 

In Fig (\ref{plot_1}) we see that  $\av{\frac{1}{l(x,x')}}$ is well
within the  bounds $b_l$ and $b_u$.  In Fig
(\ref{plot_2})  we show the percentage errors defined by 
\begin{eqnarray}
\delta_l:=\frac{1}{b_l}\bigg(\bigg\langle\frac{1}{l(x,x')}\bigg\rangle-b_l\bigg)\times 100\quad\text{and}\quad\delta_u:=\frac{1}{b_u}\bigg(\bigg\langle\frac{1}{l(x,x')}\bigg\rangle-b_u\bigg)\times 100\nonumber
\end{eqnarray}
with respect to the lower and upper bounds.  While there is a
convergence for large $N$ the error does not go to zero for either of
the bounds.  

It is also  useful to compare $\av{\frac{1}{l(x,x')}}$ to
$\frac{1}{\av{l(x,x')}}$ since it is the theoretical bound on the latter which we
are using. As shown in (Fig (\ref{plot_3})) we find an almost perfect
matching of $\av{\frac{1}{l(x,x')}}$ with $\frac{1}{\av{l(x,x')}}$
even at relatively small $N$ values. We plot the percentage error in Fig
(\ref{plot_4}) where 
\begin{equation}
\Delta:=\bigg(\bigg\langle\frac{1}{l(x,x')}\bigg\rangle\bigg)^{-1}\bigg(\bigg\langle\frac{1}{l(x,x')}\bigg\rangle-\frac{1}{\av{l(x,x')}}\bigg)\times 100\nonumber
\end{equation}
which is already very small for $N \sim 200$ and dies down further as
$N$ grows. 

Using the ``FindFit'' function in Mathematica we 
find that the best fit value for  $m_3$  is in fact 1.854 for the range
of $N$ that we have considered. As can be seen in Figure
(\ref{plot_5}) the errors for this fit  are  very small. 

 \begin{figure}[h]   
\begin{subfigure}[b]{1\linewidth} 
    \caption{$\protect\av{\frac{1}{l(x,x')}}\,\text{vs}\,N$} 
    \centering 
    \includegraphics[width=0.8\linewidth]{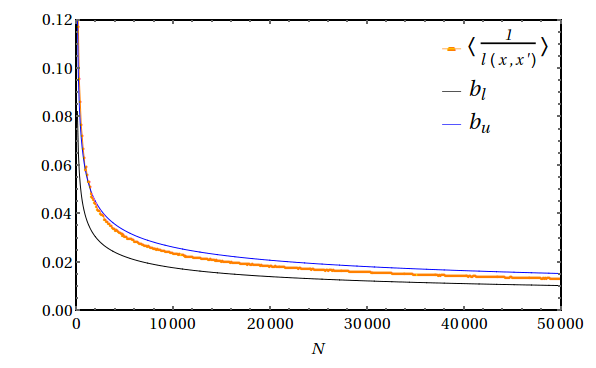}  
    \label{plot_1}
  \end{subfigure}
  \caption{Fig (a) shows a comparison of $\av{\frac{1}{l(x,x')}}$ as a
  function of $N$, with the conjectured upper and lower bounds. }
\end{figure}

\begin{figure}
\ContinuedFloat
  \begin{subfigure}[b]{1\linewidth} 
    \caption{Errors in $\protect\av{\frac{1}{l(x,x')}}$ with respect to $\protect b_u\text{, }b_l$}   
    \centering
    \includegraphics[width=0.8\linewidth]{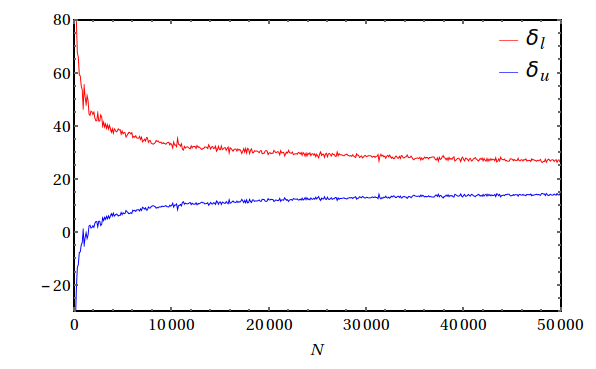} 
    \label{plot_2}
  \end{subfigure}
  \begin{subfigure}[b]{1\linewidth}
  \caption{Comparison of $\protect\av{\frac{1}{l(x,x')}}$ and $\protect\frac{1}{\av{l(x,x')}}$} 
    \centering
    \includegraphics[width=0.8\linewidth]{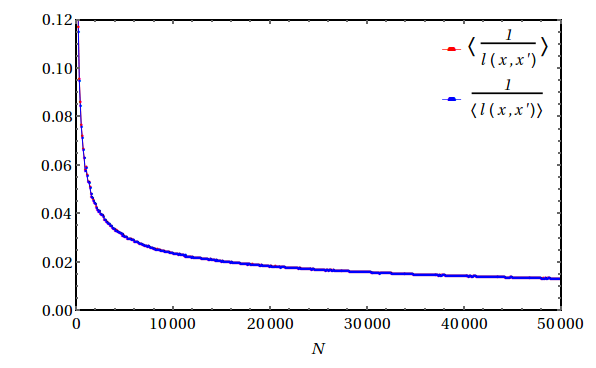}  
    \label{plot_3} 
  \end{subfigure}
  \caption{Fig
  (b) gives the  percentage error estimation with respect to these
  bounds. Fig (c) shows $\protect\av{\frac{1}{l(x,x')}}$ and
    $\protect\frac{1}{\av{l(x,x')}}$ vs $N$}
 \end{figure}

  \begin{figure}
  \ContinuedFloat
  \begin{subfigure}[b]{1\linewidth}
  \caption{Error in $\protect\frac{1}{\av{l(x,x')}}$ with respect to $\protect\av{\frac{1}{l(x,x')}}$}
    \centering
    \includegraphics[width=0.8\linewidth]{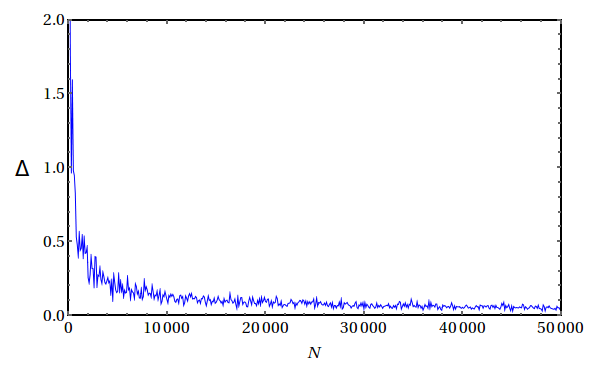} 
    \label{plot_4} 
  \end{subfigure} 
  \begin{subfigure}[b]{1\linewidth}
  \caption{Error in $\protect\av{\frac{1}{l(x,x')}}$ with respect to the Best Fit}
    \centering
    \includegraphics[width=0.8\linewidth]{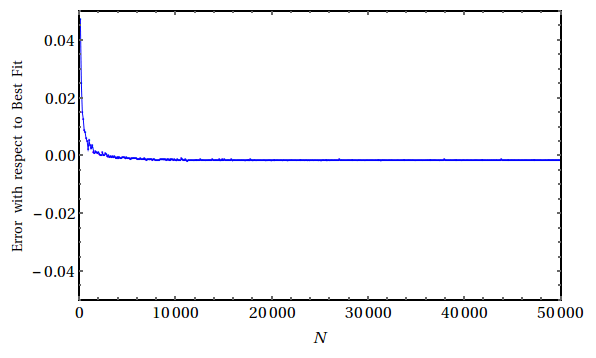} 
    \label{plot_5} 
  \end{subfigure}
  \caption{Fig (d) shows the
    percentage error between $\protect\frac{1}{\av{l(x,x')}}$ and $\protect\av{\frac{1}{l(x,x')}}$. This rapidly goes to zero as $N$
    increases. Fig (e) shows the difference between $\protect\av{\frac{1}{l(x,x')}}$ and the best fit with $\protect m_3=1.854$, this too goes to zero rapidly.}
\end{figure}

\end{document}